\documentclass[11pt]{article}

\usepackage[preprint,nonatbib]{nips_gibs}

\usepackage{amsmath,amssymb}
\usepackage{url} 

\usepackage[round,colon,authoryear,sectionbib]{natbib} 
\usepackage{amsthm}

\usepackage{enumitem}
\setitemize{noitemsep,topsep=0pt,parsep=2pt,partopsep=2pt}
\setenumerate{noitemsep,topsep=8pt,parsep=8pt,partopsep=8pt}

\newcommand{\dPDM} {\normalfont{\textsf{dPDM}}}

\newcommand{\calB}{\ensuremath{\mathcal{B}}}
\newcommand{\calN}{\ensuremath{\mathcal{N}}}
\newcommand{\calS}{\ensuremath{\mathcal{S}}}

\newtheorem{theorem}{Theorem}

\begin{document}

\title{Fair and Decentralized Exchange of Digital Goods}

\author{Ariel Futoransky \\
	Disarmista \\
	\emph{\small futo@disarmista.com} 
	\And Carlos Sarraute \\
	Wibson \& Grandata \\
	\emph{\small charles@grandata.com}
	\And Daniel Fernandez \\
	Wibson \\
	\emph{\small daniel@wibson.org}
	\AND Matias Travizano \\
	Wibson \\
	\emph{\small mat@wibson.org}
	\And Ariel Waissbein \\
	Disarmista \\
	\emph{\small wata@disarmista.com}
}

\maketitle

\begin{abstract}
We construct a privacy-preserving, distributed and decentralized marketplace 
where parties can exchange data for tokens. 
In this market, buyers and sellers make transactions in a blockchain
and interact with a third party, called notary, who has the 
ability to vouch for the authenticity and integrity of the data. 
We introduce a protocol for the data-token exchange
where neither party gains more information than what it is paying for, 
and the exchange is fair: either both parties gets the other's item or 
neither does. 
No third party involvement is required after setup, and 
no dispute resolution is needed. 
\end{abstract}

\section{Introduction}
Fair exchange, smart contracts and contingent payments have occupied 
the cryptographic community for decades. They  relate to 
problems where parties exchange a piece of data or a service, for 
data or payment. We look here upon the problem of selling personal data.

The importance of privacy in digital transactions continues to grow in 
these times, when marketing is ubiquitous and based on the footprint of
digitalized lives. 
Individuals are willing to loosen their privacy expectations in exchange 
for a service (e.g., social networks or webmail services) or even 
money. 

Consider a a distributed and decentralized marketplace where 
some parties can buy personal data. 
For example, an apparel company is willing to pay for the 2-weeks 
browsing history of people who went to Coachella Festival,
or a financial institution willing to pay for the last places 
visited by people between the ages of 20 and 35, earning over \$150K, 
in New York City.
We are interested in (i) a question that can be 
encoded in a predicate that we call audience matching criteria, 
and (ii) a payload, or piece of information, tied to this predicate. 

In this market, a {\em buyer} publishes an offer that includes
an audience match criterion, a payload description, and an amount 
he is willing to pay for an answer.
A trusted third party creates certificates for the sellers,
wherein a certificate contains an authenticated answer for an offer 
(e.g., authenticated audience match value and payload).
Each seller receives certificates, and when he receives an offer, 
he interacts with the buyer to evaluate the audience criterion. 
When there is a match, the buyer starts a contract in a blockchain
and the seller closes the contract.
With an atomic swap, tokens are exchanged for the data.
This trusted third party can be materialized as a bank, 
who can certify questions about spending and financial status;
or a telecommunications company, which can certify 
the geographical location of its subscribers (via triangulation
with the antennas their cell phones connect to).

\subsection*{Contributions and Related Work}

We introduce a decentralized and distributed marketplace for selling
personal digital data.
The  market design we present was used as a blueprint for a 
real construct, and the abstractions adhere to realistic security
problems.

We design an {\sf UC}-ideal functionality which implements secure
exchanges, and a real life protocol that securely realizes this.
This protocol is privacy friendly: it hides data and transactions
from the public.
If parties are honest, then the buyer, and only the buyer, can access 
this data and the seller gets paid. 

The protocol and marketplace formalize an existing data marketplace 
named Wibson~\citep{travizano2018wibson,fernandez2020wibson}. 
The secure exchange mechanism was presented in~\cite{futoransky2019secure}.
This marketplace uses an Ethereum side-chain, 
and a gas efficient protocol named BatPay for the recurrent micropayment of tokens~\citep{batpay2020}.
Users connect through a mobile app in their phones, and check
a message board where buyers post their offers.
The solution herein needs to conforms with a high standard both
in terms of the security and privacy, and in terms of the 
computational and communication costs.
Additionally, this marketplace features a cryptographic primitive called WibsonTree, designed to preserve users' privacy by allowing them to demonstrate predicates on their personal attributes, without revealing the values of those attributes~\citep{wibsontree2020}.

In a fair exchange, two parties wish to exchange a piece of data each holds
and is secret to the other. 
While a result by \cite{cleve1986limits} shows that basic
fair exchange is impossible in a 2-party setting, modifications
and restrictions demonstrate that it is achievable 
\citep{AsokanShoupWaidner:1997,Micali:2003:SFO,OkadaManabeOkamoto:2008:SCIS,campanelli2017zero}.

The setting of the paper differs from the models underlying these
articles. \cite{campanelli2017zero} uses a blockchain to perform
atomic swaps, but moves to use this for providing services, such
as Sudoku-puzzle solving through ZK-proofs. 
%

\section{A Model for Decentralized Exchange of Digital Goods}

We use the Universal Composable Security framework 
\citep{Canetti:2001:UCS} to formalize 
the notion of a marketplace. 
We use the $\mathcal{F}_\mathtt{ca}$--hybrid model, where parties 
are allowed to register and retrieve public keys from a certification 
authority \citep{Canetti:2003:239}.
We use the plain communication model with unauthenticated 
asynchronous communication, without guaranteed delivery 
and with possible replay (see for example \cite{Canetti:2018:067}).
Furthermore, we restrict attacks to static Byzantine 
corruptions.

The protocol is played by two main parties, a seller $\mathcal{S}$ 
and a buyer $\mathcal{B}$. 
A third party, the notary $\mathcal{N}$, vouches for the authenticity 
of data. 
That is, the notary may provide a seller with a `certificate'
which includes  data and an audience match value  
{\em binded} to the seller's id.
A fourth party, $\mathcal{W}$, plays the role of a blockchain. 
It is known \citep{BadertscherMaurerTschudiZikas:2017:149,GarayKiayiasLeonardos:2015,KiayiasZhouZikas:2015:574}
that not all the properties of a blockchain are captured by 
an Interactive Turing Machine (ITM).
Nonetheless, we model the blockchain as an ITM for the sake 
of simplicity, capturing some of its properties. 

Parties have an account with the blockchain that is binded 
to their $id$, i.e., they registered a signature verification 
public key $v$ with $\mathcal{F}_\mathtt{ca}$ and $\mathcal{W}$ 
has retrieved them all.
Furthermore, we assume that the notary $\mathcal{N}$ is 
known by all parties, i.e., they know that the id $\mathcal{N}$
is associated with a notary that can sign data and this data
ought to be trusted.
$\mathcal{W}$ stores a table (ledger) where each party id is 
associated with an amount of tokens 
the party owns, and publishes updates on each output. 
Assume that on initialization, $\mathcal{W}$ receives the 
initial state of the ledger.
At any point $\mathcal{W}$ is allowed to make token transfers
from  one account to the other, or to momentarily immobilize
a token. We assume no tokens are created or removed.
In order to model the blockchain's ability to multicast messages,
we assume that $\mathcal{W}$ writes messages to a `public' tape,
and the adversary may decide deliver it to any party, like it
does with any message.
Similarly, we assume that buyers multicast messages
and the adversary may decide what to do with them.

The blockchain, $\mathcal{W}$, executes a contract: it 
receives messages of the form 
$$
(\mathsf{Contract\,Open},id,\text{`Pay N if }x:\mathsf{H}(x)=X\text{'})
$$
and 
$$(\mathsf{Contract\,Close},sid,K).$$ 
When $\mathcal{W}$ receives a $\mathsf{Contract\,Open}$--message, 
if this is the first contact with this $id$ and the sender has
N tokens for payment, then $\mathcal{W}$ immobilizes N tokens, 
and puts the message in its public tape. 
When receiving a $\mathsf{Contract\,Close}$--message,  $\mathcal{W}$
checks for a stored $\mathsf{Contract\,Open}$--message $sid$ 
which has not been closed, he checks if $\mathsf{H}(K)=X$. 
If equality holds, $\mathcal{W}$ transfers N tokens from sender 
of the Open message to sender of the Close message, and considers
the contract closed. 
Else, $\mathcal{W}$ ignores the $\mathsf{Contract\,Close}$-message 
and wait for a new one.

We assume that we are given a function $f$ defining all audience criteria. 
For a seller attribute $\mathbf{s}$ and a buyer 
attribute $\mathbf{b}$, $f$ can be evaluated in polynomial
time and $f(\mathbf{s},\mathbf{b})=1$ if, and only if, $\mathbf{s}$
matches the criterion defined by $\mathbf{b}$.
We assume that all the negotiation details are encoded in 
$\mathbf{s}$ and $\mathbf{b}$, including the price offered 
by the buyer, which for the sake of simplicity, we have fixed
at 1 token.
%
\subsection*{Preliminaries}

Let $\Pi=(\mathsf{Gen},\mathsf{Enc},\mathsf{Dec})$ be an encryption scheme
and $\mathsf{H}$ a collision-resistant hash function.
We require that  $\Pi$ offers semantic security and remains safe 
even if an adversary is given $\mathsf{Hash}(K)$ (cf. \cite{futoransky2006foundations}).
This standard requirement in practice does not follow from 
the definitions.
Nonetheless, this can be attained for example starting from 
a secure block cipher $(\mathsf{E},\mathsf{D})$ \citep{BonehShoup:2020},
a mode of operation that allows transmitting messages of polinomially-bounded 
length (with a resulting semantically-secure symmetric cipher),
and with $\mathsf{H}:\{0,1\}^n\to\{0,1\}^k$ the
(preimage-resistant) hash function $H$ constructed from 
$\mathsf{Enc}$ (see \cite{BlackRogawayShrimpton:2002}).

Let $\mathcal{F}_\mathtt{smt}$ be the ideal-process secure message 
transmission functionality of \cite{Canetti:2001:UCS} and
let $\pi_E$ be a real-life protocol securely realizing 
$\mathcal{F}_\mathtt{smt}$, e.g., the protocol of  {\em op. cit.}
constructed out of  a public-key encryption scheme that 
is semantically secure against chosen plaintext attack.
We assume the leakage function for $\mathcal{F}_\mathtt{smt}$
to provide the adversary with the message header and the
size of the secret (non-header) portion of the message.
Let $\mathcal{F}_\mathtt{sig}$ be the signature ideal
functionality  of \cite{Canetti:2003:239} and let 
$\pi_S$ be a signature scheme which is eu-cma; hence this 
protocol securely realizes $\mathcal{F}_\mathtt{sig}$ 
\citep{GoldwasserMicaliRivest:1988,Canetti:2003:239}).

\section{The Secure Exchange Functionality}\label{sec:ideal}

We introduce the (abstract) ideal functionality that enables 
secure exchanges.
The protocol has three steps, during {\bf setup} the
notary receives an input $(\mathcal{S},M,\mathbf{s})$ consisting in 
the $id$ of a seller $\mathcal{S}$, a piece of data (or message) $M$, 
and the seller's attribute $\mathbf{s}$;
next comes the {\bf offer} where the buyer advertises an audience 
criterion through an attribute $\mathbf{b}$; and finally, the 
seller responds to the offer and the {\bf exchange} takes place.

\begin{figure}[ht]
\noindent\fbox{\begin{minipage}{.97\textwidth} 
\begin{center}
\medskip \medskip
{\bf Functionality $\mathcal{F}_\mathtt{se}$}
\end{center} {\vskip -3mm}
\begin{tabular}{*{1}{p{0.95\textwidth}}}
{ \small
\begin{enumerate}
\item
Upon receiving $(\mathsf{Certify},sid,\mathcal{S},M,\mathbf{s})$
from $\mathcal{N}$ check that $sid=(\mathcal{N},sid^\prime)$ and that no 
other $\mathsf{Certify}$ message was received with the same $sid$. 

\begin{itemize}
\item
If $\mathcal{S}$ is uncorrupted, send
$(\mathsf{Cert},sid,\mathcal{S},|M|,|\mathbf{s}|)$ to the adversary.
Upon receiving $(\mathsf{Cert\, Received},sid)$ from the adversary, send 
$(\mathsf{Cert\, Received},sid)$ to $\mathcal{S}$ and store
$(sid,\mathcal{S},M,\mathbf{s})$.

\item
If the seller $\mathcal{S}$ has been corrupted, send 
$(\mathsf{Cert},sid,\mathcal{S},M,\mathbf{s})$ to the adversary
and store $(sid,\mathcal{S},M,\mathbf{s})$.
\end{itemize}

\item
Upon receiving an offer $(\mathsf{Buy},bid, \mathbf{b})$ from 
a buyer, forward to the adversary. Each time the 
adversary sends a message  $(\mathsf{Buy},bid, \mathbf{b}, id)$ 
send $(\mathsf{Buying},bid, \mathbf{b})$ to party with id $id$ and
store the message. 

\item
Upon receiving $(\mathsf{Seller},bid,sid)$, retrieve (if any) 
a stored certificate with id $sid$, and an offer with id $bid$. 
If $f_\mathbf{s}(\mathbf{b})=1$, send the message 
$(\mathsf{Seller}, bid, sid)$ to the adversary.
\begin{itemize}
\item 
Upon receiving $(\mathsf{Contract\,Close},bid)$ from the adversary, 
transfer 1 token from $\mathcal{B}$ to $\mathcal{S}$.
\item
Upon receiving $(\mathsf{Finished},bid)$ from the adversary when
the $\mathcal{N}$ is uncorrupted, send $(\mathsf{Message},bid,M)$ 
to $\mathcal{B}$, or 
upon receiving $(\mathsf{Finished},bid,M^\prime)$ from the adversary
when $\mathcal{N}$ is corrupted, send $(\mathsf{Message},bid,M^\prime)$ 
to $\mathcal{B}$; then abort.
\end{itemize}
\end{enumerate}
}
\end{tabular}
\end{minipage}}
\caption{The ideal functionality $\mathcal{F}_\mathtt{se}$.}
\label{fig:ideal}
\end{figure}

The ideal functionality shown in Fig.~\ref{fig:ideal} gives the adversary the power
to decide who gets the offers, as the adversary controls 
the communication network, and hence also if the seller's
message, which closes the transaction, reaches the blockchain,
and if the blockchain's message can reach $\mathcal{B}$ or
$\mathcal{S}$.

%
\section{The Secure Exchange Protocol}\label{sec:real}

Let $k\in\mathbb{Z}$.
The protocol $\rho_\mathtt{se}$ in the 
$(\mathcal{F}_\mathtt{smt},\mathcal{F}_\mathtt{sig},\mathcal{F}_\mathtt{ca})$-hybrid
model goes as follows.
\begin{enumerate}
\item
When the notary $\mathcal{N}$ receives an input
$(\mathsf{Certify},sid,\mathcal{S},M,\mathbf{s})$,
he computes $K\leftarrow\mathsf{Gen}(1^k)$,
the ciphertext $C\leftarrow\mathsf{Enc}(K,M)$, the hashes 
$Y:=\mathsf{H}(C),X:=\mathsf{H}(K)$, 
and invokes $\mathcal{F}_\mathtt{sig}$ with
$(\mathsf{Sign},sid,(\mathbf{s},Y,X))$ to receive
a signature $\sigma$. 
When done, $\mathcal{N}$ sends 
$(\mathsf{Cert},sid,K,M,C,\mathbf{s},Y,X,\sigma)$ 
to $\mathcal{S}$ via $\mathcal{F}_\mathtt{smt}$.

\item
Upon receiving 
$(\mathsf{Cert},sid,K,M,C,\mathbf{s},Y,X,\sigma)$, the seller 
invokes $\mathcal{F}_\mathtt{ca}$ with $(\mathsf{Retrieve},\mathcal{N})$
and waits for $(\mathsf{Retrieve},\mathcal{N},v)$, 
next he invokes $\mathcal{F}_\mathtt{sig}$ with 
$(\mathsf{Verify},$ $ \mathcal{N},(\mathbf{s},Y,X),\sigma,v)$
and waits for an answer. 
If the signature is valid, $Y=\mathsf{H}(C),X=\mathsf{H}(K)$ 
and $M=\mathsf{Dec}(K,C)$, he outputs $(\mathsf{Cert\, received},sid)$.

\item
When $\mathcal{B}$ receives $(\mathsf{Buy},bid,\mathbf{b})$, 
he multicasts $(\mathsf{Buying},bid,\mathbf{b})$.

\item
Upon receiving $(\mathsf{Buying},bid,\mathbf{b})$, 
and if it is the first offer with this $bid$, a seller 
stores the message and outputs $(\mathsf{Offer\,received},bid)$. 

\item
When the seller receives  $(\mathsf{Sell},sid,bid)$,
he looks in his storage for a certificate and an offer with these ids.
If he finds them and $f(\mathbf{b},\mathbf{s})=1$, $\mathcal{S}$
sends the message 
$(\mathsf{Selling},bid,\mathcal{N},C,\mathbf{s},Y,X,\sigma)$ to 
$\mathcal{B}$ via $\mathcal{F}_\mathtt{smt}$ and stores $(sid,bid)$.

\item
When the buyer receives a $\mathsf{Selling}$--message,
he invokes $\mathcal{F}_\mathtt{ca}$ with 
$(\mathsf{Retrieve},\mathcal{N})$ waits to receive a 
tag $v$, he invokes $\mathcal{F}_\mathtt{sig}$ with 
$(\mathsf{Verify},\mathcal{N},(\mathbf{s},Y,X),\sigma,v)$ 
waits to receive an affirmative verification, and 
if this happens checks that $f(\mathbf{b},\mathbf{s})=1$. 
If anything fails, he ignores. Else, he sends the message
$(\mathsf{Contract\,Open},bid,$ $\text{`Pay if }x:\mathsf{H}(x)=X\text{'})$
to $\mathcal{W}$.

\item
Upon receiving a $\mathsf{Contract\,Open}$--message, 
$\mathcal{W}$ checks if the sender has a balance of at least one
token and ignores of he does not. Else, $\mathcal{W}$ immobilizes a 
token from the sender and multicasts the $\mathsf{Contract\,Open}$--message.

\item
Upon reading a $\mathsf{Contract\,Open}$--message for a $bid$ 
$\mathcal{S}$ for a stored pair $(bid,sid)$, retreives the key 
associated to $sid$ and sends $(\mathsf{Contract\,Close},bid,K)$ to 
$\mathcal{W}$.

\item
Upon receiving $(\mathsf{Contract\,Close},bid,K)$ from $\mathcal{S}$, 
the blockchain $\mathcal{W}$ looks for an associated $\mathsf{Contract\,Open}$--message
and ignores if not found or if it is already closed. Else, he computes $\mathsf{H}(K)$. 
If it agrees with $X$,  then he multicasts $(\mathsf{Contract\,Close},bid,K)$ 
in his public tape and transfers the immobilized token to 
$\mathcal{S}$ and modifies the ledger to reflect this change. 

\item
Upon reading $(\mathsf{Contract\,Close},bid,K)$, $\mathcal{B}$ outputs 
$(\mathsf{Message},bid,\mathsf{Dec}(K,C))$.

\item
Upon reading in the new state of the ledger that he was payed, the seller 
outputs $(\mathsf{Payment\,received},bid)$.
\end{enumerate}

\section{The Main Result}\label{sec:theo}
We now prove that the real-life protocol securely realizes the 
ideal functionality.

\begin{theorem}\label{theo}
Protocol $\pi_\mathtt{se}$ securely realizes  
$\mathcal{F}_{\mathtt{se}}$ in the 
$\mathcal{F}_\mathtt{ca}$--hybrid model.
\end{theorem}

\begin{proof}
By the {\sf UC} composition theorem it suffices to show that
$\rho_\mathtt{se}$ securely realizes functionality 
$\mathcal{F}_{\mathtt{se}}$ in the 
$(\mathcal{F}_\mathtt{smt},\mathcal{F}_\mathtt{sig},\mathcal{F}_\mathtt{ca})$--hybrid 
model.
Let $\mathcal{A}$ be a 
$(\mathcal{F}_\mathtt{sig},\mathcal{F}_\mathtt{smt},\mathcal{F}_\mathtt{ca})$--hybrid
model adversary.
We define an ideal-process adversary (e.g., a simulator) $\widetilde{\mathcal{A}}$ 
such that the execution in the ideal-process model and in the 
$(\mathcal{F}_\mathtt{smt},\mathcal{F}_\mathtt{sig},\mathcal{F}_\mathtt{ca})$-hybrid
models are indistinguishable by any environment $\mathcal{E}$.
The simulator runs an execution with the environment $\mathcal{E}$ 
and, in parallel, simulates a virtual copy of the hybrid adversary 
$\mathcal{A}$. 
That is, $\widetilde{\mathcal{A}}$ acts as an interface between 
$\mathcal{A}$ and $\mathcal{E}$ by imitating a copy 
of a real execution of $\rho_\mathtt{se}$ for $\mathcal{A}$, 
incorporating $\mathcal{E}$'s ideal-model interactions and 
forwarding $\mathcal{A}$'s messages to 
$\mathcal{E}$.

\medskip
{\bf Assume no corruptions happens}. 

When $\widetilde{\mathcal{A}}$ 
receives $(\mathsf{Cert},sid,\mathcal{S},|M|,|\mathbf{s}|)$ from the 
ideal functionality $\mathcal{F}_\mathtt{se}$, 
$\widetilde{\mathcal{A}}$ computes: 
\begin{itemize}
\item $K\leftarrow\mathsf{Gen}(1^k)$,
\item a random string $\tilde M\in\{0,1\}^{|M|}$ of size $|M|$, 
\item $\tilde C\leftarrow\mathsf{Enc}(K,\tilde M)$,
\item the hashes
$Y:=\mathsf{H}(C),X:=\mathsf{H}(K)$,
\end{itemize}
and simulates 
$\mathcal{F}_\mathsf{sig}$ sending $(\mathsf{Sign},\mathcal{N},(\mathsf{s},Y,X))$
to $\mathcal{A}$.

When $\mathcal{A}$ answers with $(\mathsf{Signature},\mathcal{N},(\mathsf{s},Y,X),\sigma)$, 
$\widetilde{\mathcal{A}}$ simulates $\mathcal{N}$ sending
$$
(\mathsf{Cert},\!sid,\!|(K,M,C,\mathbf{s},Y,X,\!\sigma)|)
$$ to
$\mathcal{S}$ through $\mathcal{F}_\mathtt{smt}$.\footnote{Note
that $\widetilde{\mathcal{A}}$ only needs the size of the
message and not its contents, to simulate this for 
$\mathcal{A}$, e.g., he can send a string
of $1$s of the correct size and $\mathcal{A}$
cannot distinguish one from the other.}
If $\mathcal{A}$ delivers a $\mathsf{Cert}$--message, then
$\widetilde{\mathcal{A}}$ simulates $\mathcal{F}_\mathtt{ca}$ sending
$(\mathsf{Retrieve},\mathcal{N},\mathcal{S})$ to $\mathcal{A}$. 
If the hybrid adversary answers $\mathsf{ok}$, 
then $\widetilde{\mathcal{A}}$ simulates $\mathcal{F}_\mathtt{sig}$ 
sending $(\mathsf{Verify},\mathcal{N},(\mathbf{s},Y,X),\sigma,v)$ 
to $\mathcal{A}$ and mimics $\mathcal{F}_\mathtt{sig}$ 
verification algorithm. 
If the verification checks out, $\widetilde{\mathcal{A}}$
sends $(\mathsf{Cert\,Received},sid)$
to  $\mathcal{F}_\mathtt{se}$.

When $\widetilde{\mathcal{A}}$ receives $(\mathsf{Buy},sid,\mathbf{b})$ 
from functionality $\mathcal{F}_\mathtt{se}$, he simulates $\mathcal{B}$
multicasting $(\mathsf{Buying},sid,\mathbf{b})$. 
Next, each time $\mathcal{A}$ delivers the message to
a party with id $id$, $\widetilde{\mathcal{A}}$ sends
$(\mathsf{Buying},sid,\mathbf{b},id)$ to $\mathcal{F}_\mathtt{se}$.

Upon $\widetilde{\mathcal{A}}$ receiving $(\mathsf{Sell},sid,bid)$ from 
the ideal functionality, $\widetilde{\mathcal{A}}$ simulates $\mathcal{S}$ 
sending $(\mathsf{Selling},bid,C,(\mathbf{s},Y,X),\sigma)$ to $\mathcal{B}$ 
through $\mathcal{F}_\mathtt{smt}$. 
Upon $\mathcal{A}$ delivering a $\mathsf{Selling}$--message, 
the ideal-process adversary simulates $\mathcal{F}_\mathtt{ca}$ 
sending $(\mathsf{Retrieve},\mathcal{N},\mathcal{B})$
to the hybrid adversary.
Party $\widetilde{\mathcal{A}}$ waits for $\mathcal{A}$ 
to answer with $\mathsf{ok}$, then $\widetilde{\mathcal{A}}$ 
simulates $\mathcal{B}$ by sending to $\mathcal{W}$
$$
(\mathsf{Contract\,Open},bid,\text{`Pay if }x:\mathsf{H}(x)=X\text{'})
$$
and waits for $\mathcal{A}$ to deliver this
to $\mathcal{W}$.
When this happens, the ideal-process adversary checks if $\mathcal{B}$
has at least one token in its account, and ignores if there
is less. 
Else $\widetilde{\mathcal{A}}$ `immobilizes' one simulated
token, and simulates $\mathcal{W}$ multicasting the $\mathsf{Contract\,Open}$--message.
When $\mathcal{A}$ delivers this message to $\mathcal{S}$, 
the ideal-process adversary simulates $\mathcal{S}$ sending 
$(\mathsf{Contract\,Close},bid,K)$ to $\mathcal{W}$.
Upon $\mathcal{A}$ delivering a $\mathsf{Contract\,Close}$--message
with id $bid$, $\widetilde{\mathcal{A}}$ verifies that $H(K)=X$
and if this happens, $\widetilde{\mathcal{A}}$ sends 
$(\mathsf{Contract\,Close},bid)$
to $\mathcal{F}_\mathtt{se}$ and simulates $\mathcal{W}$ putting 
this message in its public tape.
Upon the adversary $\mathcal{A}$ delivers this message, 
$\widetilde{\mathcal{A}}$ sends $(\mathsf{Finished},bid)$ to the
ideal functionality $\mathcal{F}_\mathtt{se}$.

In order to conclude that the result follows in the uncorrupted 
case, first notice that the delays in message delivery are
incorporated by simulator so that all outputs are synchronized.
Second, note that the hybrid adversary learns the headers: 
\begin{itemize}
\item $(\mathsf{Cert},sid)$, 
\item $(\mathsf{Selling}, sid)$,
\item the size $|(K,M,\mathbf{s},C,X,\sigma)|$,
\item the size $|(\mathbf{s},C,X,\sigma)|$.
\end{itemize}
He also learns the full contents of 
\begin{itemize}
\item $(\mathsf{Buying},sid,\mathbf{b})$,
\item $(\mathsf{Contract\,Open},sid,\text{`Pay if }x:\mathsf{H}(x)=X\text{'})$,
\item $(\mathsf{Contract\,Close},sid,K)$, 
\item and who receives the first of these messages.
\end{itemize}
These are computationally indistinguishable
and thus the result follows.

\medskip
{\bf Assume $\mathcal{A}$ corrupts $\mathcal{N}$.}

In that case 
$\widetilde{\mathcal{A}}$ corrupts $\widetilde{\mathcal{N}}$ and
has dummy $\widetilde{\mathcal{N}}$  output what 
$\mathcal{N}$ outputs.
When $\widetilde{\mathcal{A}}$ receives 
$(\mathsf{Cert},sid,\mathcal{S},M,\mathbf{s})$ from the ideal
functionality $\mathcal{F}_\mathtt{se}$, he simulates the environment 
sending this to the simulated $\mathcal{N}$ as input. 
Moreover, $\widetilde{\mathcal{A}}$ simulates the other parties for 
$\mathcal{N}/\mathcal{A}$.
In particular, if the simulated $\mathcal{N}/\mathcal{A}$ sends a 
message 
$$
(\mathsf{Cert},sid,K^\prime,M^\prime,C^\prime,\mathbf{s}^\prime,Y^\prime,X^\prime,\sigma^\prime)
$$
through $\mathcal{F}_\mathtt{smt}$ to $\mathcal{S}$,
then the ideal-process adversary continues the simulation as in the 
uncorrupted case.

The corruption of the notary can only influence the certificate sent
to the seller, which is the only message sent by the notary.
If $\mathcal{N}/\mathcal{A}$ sends a certificate which
does not have the correct format, or the signature validation is not
passed:
$$Y^\prime\neq\mathsf{H}(C^\prime) \; \mbox{   or  } \;  X^\prime\neq\mathsf{H}(K^\prime) 
\; \mbox{  or  } \; 
C^\prime\neq\mathsf{Dec}(K^\prime,C^\prime),$$
then the seller ignores
the message.
Assume then that all these controls are passed.
Then the simulator continues as in the uncorrupted case: 
$\widetilde{\mathcal{A}}$ simulates $\mathcal{W}$ 
sending the $\mathsf{Contract\,Close}$--message to
$\mathcal{B}$.
If the adversary $\mathcal{A}$ delivers this message, 
$\widetilde{\mathcal{A}}$ sends $(\mathsf{Finished},bid,M^\prime)$ 
to the ideal functionality $\mathcal{F}_\mathtt{se}$.
Hence, both the $\mathcal{B}$ and $\widetilde{\mathcal{B}}$
pay for $M^\prime$.

\medskip
{\bf Assume the hybrid adversary corrupts $\mathcal{S}$.}

In this case the 
simulator corrupts $\widetilde{\mathcal{S}}$.
Additionally he has $\widetilde{\mathcal{S}}$ output what 
$\mathcal{S}$ outputs.
When $\widetilde{\mathcal{A}}$ receives 
$(\mathsf{Cert},sid,\mathcal{S},\mathbf{s}, M)$
from the ideal functionality $\mathcal{F}_\mathtt{se}$, he proceeds 
analogously to the uncorrupted case, changing the fake values for 
real ones. 
More precisely, $\widetilde{\mathcal{A}}$ simulates key generation,
encryption and hashing with the real values, and then invoking
$\mathcal{F}_\mathtt{sig}$ with $(\mathsf{Sign},\mathcal{N},(\mathbf{s},Y,X))$.
Upon obtaining $(\mathsf{Signature},\mathcal{N},(\mathbf{s},Y,X),\sigma)$ 
from the hybrid adversary, $\widetilde{\mathcal{A}}$ simulates sending 
$$
(\mathsf{Cert},sid,K,M,C,\mathbf{s},Y,X,\sigma)
$$ from $\mathcal{N}$
to $\mathcal{S}$ through $\mathcal{F}_\mathtt{smt}$.
(Note that here $\widetilde{\mathcal{A}}$ holds $M$ and $s$,
generates $K$, computes $X,C$ and $\sigma$ --same as $\rho_\mathtt{se}$.)
When the hybrid adversary delivers the message, 
$\widetilde{\mathcal{A}}$ sends $(\mathsf{Cert\,Received},sid)$
to the ideal functionality $\mathcal{F}_\mathtt{se}$.

Upon receiving $(\mathsf{Buy},bid,\mathbf{b})$,
$\widetilde{\mathcal{A}}$ continues like in the uncorrupted case.

Upon receiving $(\mathsf{Sell},sid,bid)$ from $ \mathcal{F}_\mathtt{se}$, 
$\widetilde{\mathcal{A}}$ simulates $\mathcal{S}$ receiving the same 
message from the environment.
If the corrupted seller, $\mathcal{S}$, sends a message
$$
(\mathsf{Selling},sid,C^\prime,\mathbf{s}^\prime,Y^\prime,X^\prime,\sigma^\prime)
$$
to $\mathcal{B}$ through $\mathcal{F}_\mathtt{smt}$, the ideal-process 
adversary simulates $\mathcal{F}_\mathtt{ca}$ sending 
$$
(\mathsf{Retrieve},\mathcal{N},\mathcal{B})
$$ 
to the hybrid adversary
and if $\mathcal{A}$ answers with $\mathsf{ok}$, then
$\widetilde{\mathcal{A}}$ simulates $\mathcal{F}_\mathtt{sig}$ sending
$(\mathsf{Verify},\mathcal{N},(\mathbf{s}^\prime,Y^\prime,X^\prime),\sigma^\prime,v)$ 
and ignores if the signature validation is not passed.
Else, $\widetilde{\mathcal{A}}$ simulates $\mathcal{B}$ by sending to $\mathcal{W}$
$$
(\mathsf{Contract\,Open},sid,\text{`Pay if }x:\mathsf{H}(x)=X\text{'}) .
$$
If the hybrid adversary delivers the message, $\widetilde{\mathcal{A}}$
simulates $\mathcal{W}$ writing this message to its public tape. 
When the hybrid adversary delivers the message to the $\mathcal{S}$,
$\widetilde{\mathcal{A}}$ waits for $\mathcal{S}$ to send 
$(\mathsf{Contract\,Close},bid,K^\prime)$ to $\mathcal{W}$.
$\widetilde{\mathcal{A}}$ checks if $K^\prime=K$ and ignores if
it does not hold.
Else, the ideal-process adversary sends $(\mathsf{Contract\,Close},sid)$ 
to $\mathcal{F}_\mathtt{se}$ and simulates $\mathcal{W}$ writing 
$(\mathsf{Contract\,Close},sid,K)$ to its public tape.
If  $\mathcal{A}$ delivers, $\widetilde{\mathcal{A}}$ 
sends $(\mathsf{Finished},sid)$ to the ideal functionality 
$\mathcal{F}_\mathtt{se}$.
Again, $\mathcal{E}$ can't distinguish the
hybrid from ideal-process protocol executions. 

The cases where the adversary corrupts $\mathcal{W}$
or multiple parties is left out for the sake of 
brevity, thus concluding the proof.

\end{proof}

\section{Conclusion}
\label{sec:variants}

Here we proposed a solution to the problem of trading real-world private information using only cryptographic protocols and a public blockchain to guarantee the fairness of transactions.
We described a protocol that we call ``Secure Exchange of Digital Goods'' \citep{futoransky2019secure} between a data buyer $\calB$ and a data seller $\calS$.
The protocol relies on a trusted third party $\calN$, which also plays the role of notary in the context of a decentralized Privacy-Preserving Data Marketplace (\dPDM) such as the Wibson Marketplace \citep{travizano2018wibson,fernandez2020wibson}.

This protocol converts the exchange of data into an atomic transaction where two things happen simultaneously:
\begin{itemize}
\item The buyer $\calB$ gets access to the data, by learning the key that enables him to decrypt $C$ (previously received encrypted data).
\item The seller $\calS$ gets paid for his data by revealing the key.
\end{itemize}

There exists some open questions that cannot be addressed
completely in this paper, but are worth posing to understand
the value of the model.
For example, can we modify the protocol so that $\mathsf{s}$ or 
$\mathsf{b}$ are not shared? 
How can we allow a seller to close a contract if he does not have 
currency (e.g., Ethereum gas) to pay for the transaction?
We have actually developed a solution for this problem in \citep{batpay2020}.

The implementation of the above protocol implies costs to all intervening 
parties.
Each transaction costs at least the computation of a hash,
and if this turns to be expensive, the protocol needs to be 
optimized. This can be done with the BatPay smart contract \citep{batpay2020}.





\bibliography{../wibson.bib}

\begin{thebibliography}{20}
\providecommand{\natexlab}[1]{#1}
\providecommand{\url}[1]{\texttt{#1}}
\expandafter\ifx\csname urlstyle\endcsname\relax
  \providecommand{\doi}[1]{doi: #1}\else
  \providecommand{\doi}{doi: \begingroup \urlstyle{rm}\Url}\fi

\bibitem[Asokan et~al.(2000)Asokan, Shoup, and
  Waidner]{AsokanShoupWaidner:1997}
Nadarajah Asokan, Victor Shoup, and Michael Waidner.
\newblock Optimistic fair exchange of digital signatures.
\newblock \emph{IEEE Journal on Selected Areas in communications}, 18\penalty0
  (4):\penalty0 593--610, 2000.

\bibitem[Badertscher et~al.(2017)Badertscher, Maurer, Tschudi, and
  Zikas]{BadertscherMaurerTschudiZikas:2017:149}
Christian Badertscher, Ueli Maurer, Daniel Tschudi, and Vassilis Zikas.
\newblock Bitcoin as a transaction ledger: A composable treatment.
\newblock Cryptology ePrint Archive, Report 2017/149, 2017.
\newblock \url{https://eprint.iacr.org/2017/149}.

\bibitem[Black et~al.(2002)Black, Rogaway, and
  Shrimpton]{BlackRogawayShrimpton:2002}
John Black, Phillip Rogaway, and Thomas Shrimpton.
\newblock Black-box analysis of the block-cipher-based hash-function
  constructions from {PGV}.
\newblock In \emph{Annual International Cryptology Conference}, pages 320--335.
  Springer, 2002.

\bibitem[Boneh and Shoup(2020)]{BonehShoup:2020}
Dan Boneh and Viktor Shoup.
\newblock \emph{A Graduate Course in Applied Cryptography}.
\newblock 2020.
\newblock Published online, see \url{https://toc.cryptobook.us/}.

\bibitem[Campanelli et~al.(2017)Campanelli, Gennaro, Goldfeder, and
  Nizzardo]{campanelli2017zero}
Matteo Campanelli, Rosario Gennaro, Steven Goldfeder, and Luca Nizzardo.
\newblock Zero-knowledge contingent payments revisited: Attacks and payments
  for services.
\newblock In \emph{Proceedings of the 2017 ACM SIGSAC Conference on Computer
  and Communications Security}, pages 229--243. ACM, 2017.

\bibitem[Canetti(2000)]{Canetti:2018:067}
Ran Canetti.
\newblock Universally composable security: A new paradigm for cryptographic
  protocols.
\newblock Cryptology ePrint Archive, Report 2000/067, 2000.
\newblock \url{https://eprint.iacr.org/2000/067}.

\bibitem[Canetti(2001)]{Canetti:2001:UCS}
Ran Canetti.
\newblock Universally composable security: A new paradigm for cryptographic
  protocols.
\newblock In \emph{Proceedings 42nd IEEE Symposium on Foundations of Computer
  Science}, pages 136--145. IEEE, 2001.

\bibitem[Canetti(2003)]{Canetti:2003:239}
Ran Canetti.
\newblock Universally composable signatures, certification and authentication.
\newblock Cryptology ePrint Archive, Report 2003/239, 2003.
\newblock \url{https://eprint.iacr.org/2003/239}.

\bibitem[Cleve(1986)]{cleve1986limits}
Richard Cleve.
\newblock Limits on the security of coin flips when half the processors are
  faulty.
\newblock In \emph{Proceedings of the eighteenth annual ACM symposium on Theory
  of computing}, pages 364--369. ACM, 1986.

\bibitem[Fernandez et~al.(2020)Fernandez, Futoransky, Ajzenman, Travizano, and
  Sarraute]{fernandez2020wibson}
Daniel Fernandez, Ariel Futoransky, Gustavo Ajzenman, Matias Travizano, and
  Carlos Sarraute.
\newblock Wibson protocol for secure data exchange and batch payments.
\newblock \emph{arXiv:2001.08832}, 2020.

\bibitem[Futoransky et~al.(2006)Futoransky, Kargieman, Sarraute, and
  Waissbein]{futoransky2006foundations}
Ariel Futoransky, Emiliano Kargieman, Carlos Sarraute, and Ariel Waissbein.
\newblock Foundations and applications for secure triggers.
\newblock \emph{ACM Transactions on Information and System Security (TISSEC)},
  9\penalty0 (1):\penalty0 94--112, 2006.

\bibitem[Futoransky et~al.(2019)Futoransky, Sarraute, Waissbein, Fernandez,
  Travizano, and Minnoni]{futoransky2019secure}
Ariel Futoransky, Carlos Sarraute, Ariel Waissbein, Daniel Fernandez, Matias
  Travizano, and Martin Minnoni.
\newblock Secure exchange of digital goods in a decentralized data marketplace.
\newblock In \emph{Proceedings of the 2019 Argentine Symposium on Big Data
  (AGRANDA)}, pages 38--44, 2019.

\bibitem[Futoransky et~al.(2020)Futoransky, Sarraute, Waissbein, Travizano, and
  Fernandez]{wibsontree2020}
Ariel Futoransky, Carlos Sarraute, Ariel Waissbein, Matias Travizano, and
  Daniel Fernandez.
\newblock {W}ibson{T}ree: Efficiently preserving seller's privacy in a
  decentralized data marketplace.
\newblock \emph{arXiv:2002.03810}, 2020.

\bibitem[Garay et~al.(2015)Garay, Kiayias, and
  Leonardos]{GarayKiayiasLeonardos:2015}
Ju\'an Garay, Agelos Kiayias, and Nikos Leonardos.
\newblock The bitcoin backbone protocol: Analysis and applications.
\newblock In E.~Oswald and M.~Fischlin, editors, \emph{Advances in Cryptology -
  EUROCRYPT 2015}, volume 9057. Springer, Berlin, Heidelberg, 2015.

\bibitem[Goldwasser et~al.(1988)Goldwasser, Micali, and
  Rivest]{GoldwasserMicaliRivest:1988}
Shafi Goldwasser, Silvio Micali, and Ronald~L. Rivest.
\newblock A digital signature scheme secure against adaptive chosen-message
  attacks.
\newblock \emph{SIAM J. Comput.}, 17\penalty0 (2):\penalty0 281--308, April
  1988.
\newblock ISSN 0097-5397.

\bibitem[Kiayias et~al.(2015)Kiayias, Zhou, and
  Zikas]{KiayiasZhouZikas:2015:574}
Aggelos Kiayias, Hong-Sheng Zhou, and Vassilis Zikas.
\newblock Fair and robust multi-party computation using a global transaction
  ledger.
\newblock Cryptology ePrint Archive, Report 2015/574, 2015.
\newblock \url{https://eprint.iacr.org/2015/574}.

\bibitem[Mayer et~al.(2020)Mayer, Bejarano, Fernandez, Ajzenman, Ayala,
  Santoalla, Sarraute, and Futoransky]{batpay2020}
Hartwig Mayer, Ismael Bejarano, Daniel Fernandez, Gustavo Ajzenman, Nicolas
  Ayala, Nahuel Santoalla, Carlos Sarraute, and Ariel Futoransky.
\newblock {B}at{P}ay: a gas efficient protocol for the recurrent micropayment
  of {ERC20} tokens.
\newblock \emph{arXiv:2002.02316}, 2020.

\bibitem[Micali(2003)]{Micali:2003:SFO}
Silvio Micali.
\newblock Simple and fast optimistic protocols for fair electronic exchange.
\newblock In \emph{Proceedings of the twenty-second annual symposium on
  Principles of distributed computing}, pages 12--19, 2003.

\bibitem[Okada et~al.(2008)Okada, Manabe, and
  Okamoto]{OkadaManabeOkamoto:2008:SCIS}
Yusuke Okada, Yoshifumi Manabe, and Tatsuaki Okamoto.
\newblock An optimistic fair exchange protocol and its security in the
  universal composability framework.
\newblock \emph{International Journal of Applied Cryptography}, 1\penalty0
  (1):\penalty0 70--77, 2008.

\bibitem[Travizano et~al.(2018)Travizano, Sarraute, Ajzenman, and
  Minnoni]{travizano2018wibson}
Matias Travizano, Carlos Sarraute, Gustavo Ajzenman, and Martin Minnoni.
\newblock Wibson: A decentralized data marketplace.
\newblock In \emph{Proceedings of SIGBPS 2018 Workshop on Blockchain and Smart
  Contract}, 2018.

\end{thebibliography}

\end{document}